\documentclass[aps,prl,showpacs,floatfix,twocolumn,amsmath,amssymb,superscriptaddress,nobalancelastpage]{revtex4-1}
\usepackage{graphicx}
\usepackage{color}
\usepackage{ulem}
\usepackage{bm}
\usepackage{hyperref}

\newcommand{\licuvo}{LiCuVO$_4$}

\newcommand{\HQ}{\ensuremath{H_{\rm Q}}}
\newcommand{\HC}{\ensuremath{H_{\rm C}}}
\newcommand{\HS}{\ensuremath{H_{\rm sat}}}
\newcommand{\kicy}{\ensuremath{k_{\rm IC}}}
\newcommand{\tikicy}{\ensuremath{\tilde{k}_{\rm IC}}}
\newcommand{\Hpc}{\ensuremath{{\bf H}\!\parallel\!\bold{c}}}
\newcommand{\Hpa}{\ensuremath{{\bf H}\!\parallel\!\bold{a}}}

\newcommand{\Hpz}{\ensuremath{{\bf H}\!\parallel\!\bold{z}}}
\newcommand{\kic}{\ensuremath{{\boldsymbol{k}_{\rm IC}}}}

\def\etal{\textit{et al.}}
\def\Ang{\AA$^{-1}$}

\def\ie{{i.e.}}

\def\epl{Europhys.\ Lett.\ }

\def\inch{Inorg.\ Chem.\ }

\def\jetpl{JETP Lett.\ }

\def\jpcm{J.\ Phys.:\ Condens.\ Matter }

\def\jpsj{J.\ Phys.\ Soc.\ Jpn.\ }

\def\prb{Phys.\ Rev.\ B }
\def\prl{Phys.\ Rev.\ Lett.\ }

\def\ptps{Prog.\ Theor.\ Phys.\ Suppl.\ }

\begin{document}
\title{Evidence of a bond-nematic phase in LiCuVO$_4$}
\author{M. Mourigal}
\affiliation{Institut Laue-Langevin, BP 156, F-38042 Grenoble Cedex 9,
             France}
\affiliation{Institute for Quantum Matter 
             and Department of Physics and Astronomy,
             Johns Hopkins University, Baltimore, MD 21218, USA}
\author{M. Enderle}
\affiliation{Institut Laue-Langevin, BP 156, F-38042 Grenoble Cedex 9,
             France}
\author{B. F\aa k}
\affiliation{SPSMS, UMR-E CEA / UJF-Grenoble 1, INAC, F-38054 Grenoble, France}
\author{R. K. Kremer}
\affiliation{Max-Planck Institute for Solid State Research,
             Heisenbergstrasse 1, D-70569 Stuttgart, Germany}
\author{J. M. Law}
\affiliation{Max-Planck Institute for Solid State Research,
             Heisenbergstrasse 1, D-70569 Stuttgart, Germany}
\author{A. Schneidewind}
\affiliation{Gemeinsame Forschergruppe Helmholtz-Zentrum Berlin -- TU Dresden, 
		   D-85747 Garching, Germany}
\author{A. Hiess}
\thanks{present address: European Spallation Source, ESS AB, P.P. Box 176, SE-22100 Lund, Sweden.}
\affiliation{Institut Laue-Langevin, BP 156, F-38042 Grenoble Cedex 9,
             France}
\author{A. Prokofiev}
\affiliation{Johann Wolfgang von Goethe Universit\"at, Postfach 111932, 
             D-60054 Frankfurt, Germany} 
\affiliation{Institute of Solid State Physics, Vienna University of 
		   Technology, A-1040 Vienna, Austria}

\date{\today}
\begin{abstract}
Polarized and unpolarized neutron scattering experiments on the frustrated ferromagnetic spin-1/2 chain \licuvo\ show that the phase transition at \HQ\ of 8~Tesla is driven by quadrupolar fluctuations and that dipolar correlations are short-range with moments parallel to the applied magnetic field in the high-field phase. Heat-capacity measurements evidence a phase transition into this high-field phase, with an anomaly clearly different from that at low magnetic fields. Our experimental data are consistent with a picture where the ground state above \HQ\ has a next-nearest neighbour bond-nematic order along the chains with a fluid-like coherence between weakly coupled chains.
\end{abstract}
\maketitle

Frustrated quantum ferromagnets are predicted to display a new exotic state of matter called a bond-nematic state, a highly correlated quantum spin state resembling nematic liquid crystals~\cite{Momoi05,Shannon06,Vekua07}. Starting from a magnetically saturated high-field state, this exotic phase arises from Bose-Einstein condensation of two-magnon pairs upon lowering the magnetic field. Unlike conventional Bose-Einstein condensates, there is no transverse long-range dipolar order but (quasi-) long-range quadrupolar-nematic order. The corresponding order parameter is defined on the bond between two neighboring sites by a rank-2 tensor product of the local spin degrees of freedom. This new type of quantum ground state  should be observable in real materials, since it is predicted to exist for a large variety of 1D~\cite{Vekua07,Kecke07,Hikihara08,Heidrich09,Sudan09}, 2D~\cite{Momoi05,Shannon06,Shindou09,Zhitomirsky10} and even some 3D spin-1/2 ferromagnets~\cite{Ueda09}, which all have in common a strong next-nearest-neighbor frustrating exchange. A particularly interesting model is the spin-1/2 frustrated ferromagnetic Heisenberg (FFH) chain, where a quadrupolar-nematic phase is predicted to extend from low magnetic fields to the saturation field \HS, with algebraic decay of transverse quadrupolar and longitudinal dipolar correlations  \cite{Vekua07,Kecke07,Hikihara08,Sato09}. 
In view of the strong theoretical activity in this field, clearly identified experimental realizations of bond-nematic states are needed.

The FFH chain \licuvo\ is an excellent candidate to search for an experimental realization of bond-nematic correlations. The excitation spectrum displays pronounced quantum effects~\cite{Enderle05,Enderle10} and a  magnetization anomaly near $\mu_0\HC\!\sim\! 40$--$48$~T~\cite{Svistov11} corresponds to a predicted long-range ordered quadrupolar bond-nematic phase~\cite{Zhitomirsky10}. Between \HC\ and an additional magnetization anomaly at $\mu_0\HQ\!\sim\!8$~T~\cite{Banks07}, the order has been supposed to be dipolar long-ranged and amplitude-modulated~\cite{Buettgen07,Masuda11}. However, the character of the ordered phases above \HQ\ remains a subject of speculation.

In this Letter we present experimental evidence that in \licuvo\, the field-induced phase transition at \HQ\ is driven by quadrupolar correlations. Based on neutron scattering, we show explicitly that the dipolar correlations, which are long-range (LR) at zero field due to a weak interchain coupling, are in fact short-range (SR) in all crystallographic directions above \HQ\ down to the lowest temperatures (100 mK) and involve only the spin components parallel to the magnetic field. The field-dependent characteristic wave vector follows precisely the theoretical predictions for the quadrupolar-nematic phase, namely $\kicy\!=(1/2-m)/p$, where $m\!=\!\langle S_n^z\rangle$ is the uniform  magnetization and $p\!=\!2$ for the quadrupolar bond-nematic state~\cite{Vekua07,Kecke07,Hikihara08,Sudan09}. A concomitant change in order parameter (or universality class) is revealed by heat capacity measurements.

The Cu$^{2+}$ ions in \licuvo\ form spin-1/2 chains along the orthorhombic $b$ axis~\cite{Gibson04,Koo11}. Nearest-neighbor (NN) spins are coupled ferromagnetically (FM) ($J_1\!<\!0$) and next-nearest neighbors (NNN) antiferromagnetically (AF) ($J_2\!>\!0$)~\cite{Enderle05}. Small interchain interactions lead to 3D LR order below $T_N\!\approx\!2.4$~K, where the reduced ordered moments ($m_0\!\approx\!0.3\,\mu_{B}$) form a circular cycloid in the $ab$-plane with incommensurate propagation vector $\kic\!=\!(0,0.532,0)$~\cite{Gibson04,Mourigal11}, which corresponds to $\kicy=0.468$~r.l.u. for a 1D chain due to the centering translation of the $Imma$ space group and $\tikicy=0.234$ in units of $2\pi/d_{\rm NN}$, where $d_{\rm NN} = b/2$ is the NN distance.

\begin{figure}
\includegraphics[width=0.99\columnwidth]{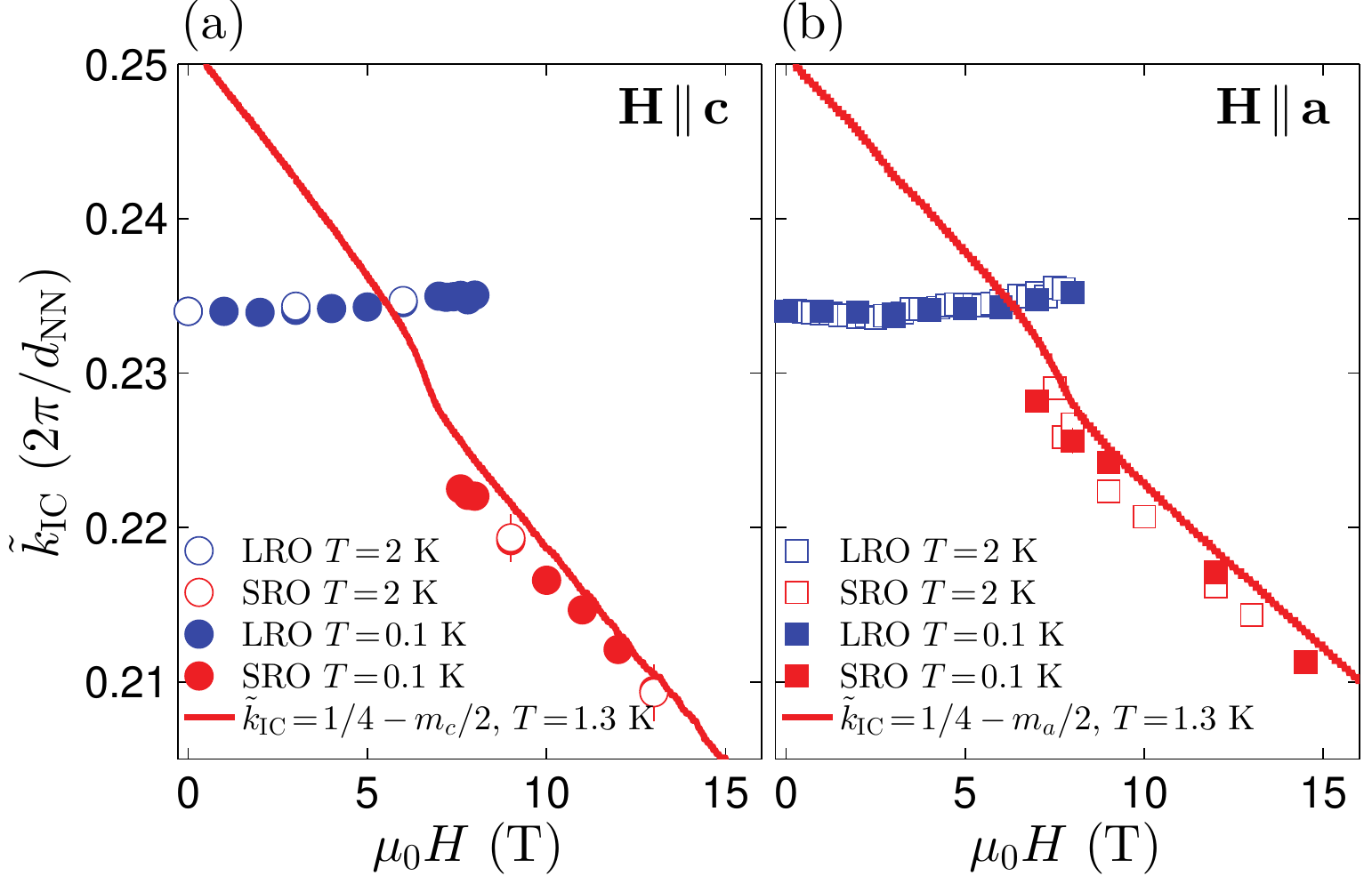}
\includegraphics[width=0.99\columnwidth]{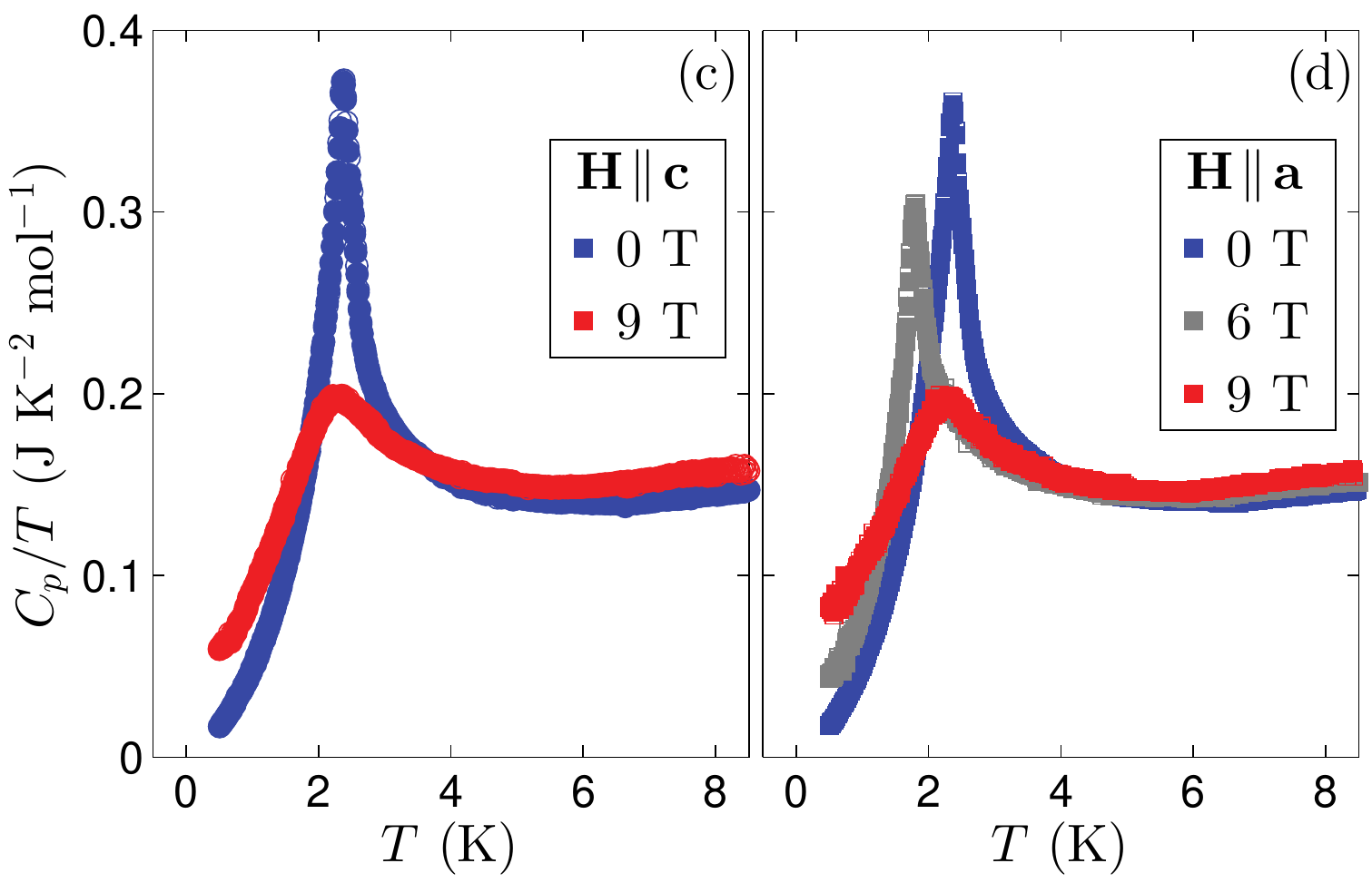}
\caption{(Color online) 
(a--b) Incommensurate propagation vector $\tikicy$ in units of $2\pi/d_{\rm NN}$ as a function of applied field  for \Hpc\ (left) and \Hpa\ (right) compared to the theoretical prediction for the quadrupolar-nematic phase \cite{Vekua07,Kecke07,Hikihara08,Sudan09}, using the experimental magnetization $m_{a,c}$ of Ref.~\cite{Svistov11}, normalized to 1/2 at the saturation field. (c--d) Specific heat for \Hpc\ (left) and \Hpa\ (right). The distinct anomaly evidences a phase transition at all fields and the change in shape near $\HQ\!\sim\!8$~T suggests a change of the universality class.}
\label{FigKC}
\end{figure}

A first series of unpolarized elastic neutron scattering measurements were performed on the triple-axis spectrometers PANDA at FRM~II (Munich) and IN14 and IN20 at Institut Laue-Langevin (Grenoble), with vertically focusing monochromator and analyzer and incident wave vectors $1.5\!\le\!k_i\!\le\!2.662$~\Ang. Second order contamination was suppressed by a cooled Be-filter or a pyrolythic graphite filter. The same single crystal as used in Ref.~\cite{Enderle10} was mounted with either the $a$- or the $c$-axis along the vertical field of a 15~T split-coil cryomagnet. A dilution insert cooled the sample to 100~mK.

Figure~\ref{FigKC}(a,b) shows the magnetic field dependence of the propagation vector of dipolar spin correlations within the chain, \tikicy. Below $\mu_0\HQ\!\sim\!8$~T, \tikicy\ is essentially field-independent while above \HQ\ it follows the relation $\tikicy\!=(1/2-m_{a,c}(H))/p$ with $p\!=\!2$, where $m_a$ and $m_c$ are the magnetization curves measured at 1.3~K \cite{Svistov11}, normalized to $m=1/2$ at the saturation field. Both below and above \HQ, \tikicy\ behaves as predicted theoretically \cite{Vekua07,Kecke07,Hikihara08,Sudan09}. Above \HQ\, the slope of \tikicy\ versus $m$, \ie\ $p=2$, proves the presence of quadrupolar correlations for both \Hpc\ [Fig.~\ref{FigKC}(a)] and \Hpa\ [Fig.~\ref{FigKC}(b)]. 

\begin{figure}
\includegraphics[width=0.99\columnwidth]{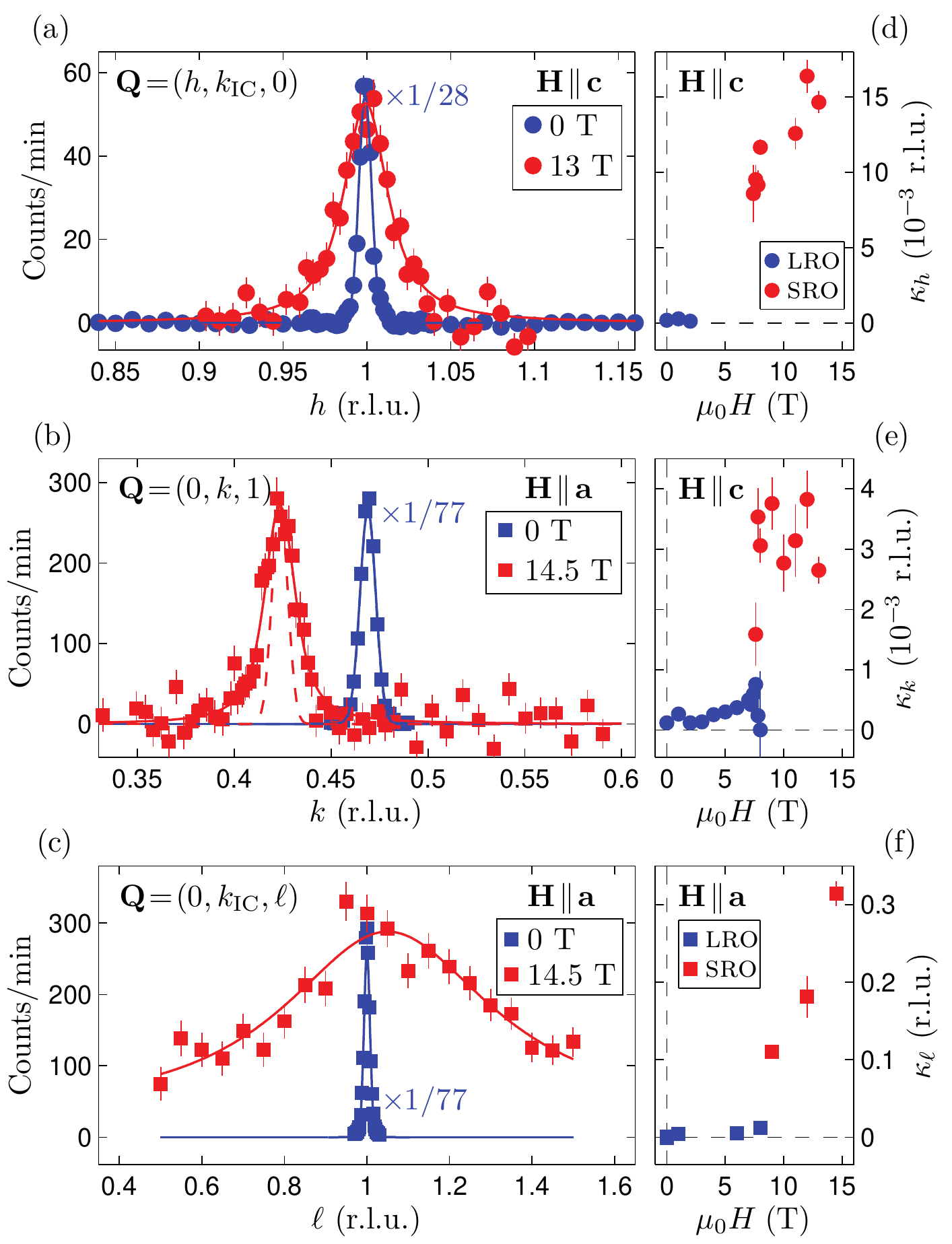}
\caption{(Color online) (a--c) Scans of the dipolar correlations at $T\!=\!0.1$~K for the three reciprocal space directions are resolution limited at zero field and broadened at high fields above $\HQ\!\sim\!8$~T. (d--f) Magnetic field dependence of the inverse dipolar correlation length $\kappa=\xi^{-1}$. The onset of short-range order at \HQ\ is clearly seen for all three reciprocal space directions.}
\label{FigWidth}
\end{figure}

The specific heat close to the phase transition from the paramagnetic state into the magnetically ordered phases was measured employing the relaxation method using a Quantum Design calorimeter. Despite an obvious change in the peak shape, see Fig.~\ref{FigKC}(c,d), the anomalies at 9~T for \Hpa\ and \Hpc\ remain remarkably well-defined and clearly different from an SR order maximum, thus testifying a phase transition. The pronounced change of the shape of the specific heat peak above \HQ\ indicates a different universality class compared to lower fields. The peak shape above \HQ\ is identical for \Hpc\ and \Hpa, evidencing the same universality class for the two field directions.

Below \HQ, the order parameter is dipolar as shown by the sharp (resolution-limited) magnetic Bragg peaks in our neutron scattering measurements. Above \HQ, these peaks are clearly broadened for all three crystallographic directions, as shown in Fig.~\ref{FigWidth}(a--c). The increase in the peak width appears abruptly at \HQ, as evidenced in Fig.~\ref{FigWidth}(d--f), where the intrinsic peak width $\kappa$, which is inversely proportional to the dipolar correlation length $\xi$, is plotted as a function of applied field for the three space directions. The occurrence of a {\it finite} dipolar correlation length $\xi$ above \HQ\ clearly establishes that the order parameter is no longer dipolar. 
In order to rule out the hypothetical existence of a dipolar LR order with a different wave vector, we have estimated the integrated intensity involved in the SR dipolar correlations. We first note that both \kicy\ and $\kappa_k$ are $\ell$-independent, since scans along $(0,k,1.5)$ and $(0,k,1)$ for \Hpa\ yield identical \kicy\ and $\kappa_k$. The integrated intensity above \HQ\ can then be obtained by explicit integration over $\ell$ in $(0,\kicy,\ell)$~[Fig.~\ref{FigWidth}(c)] and $k$ in $(0,k,1)$~[Fig.~\ref{FigWidth}(b)], while the $h$-integration is effectively achieved by the coarse vertical resolution in the measurements. The estimated integrated intensity at 14.5~T is at least as high as at zero field, \ie, the dipolar SR correlations at high field involve at least as much of the magnetic moment as the LR order at zero field. We can therefore exclude a conventional dipolar LR order with another ordering wave vector. The disparate correlation lengths of the SR dipolar correlations along $a$, $b$, and $c$ may be related to the hierarchy of the exchange interactions in \licuvo\ \cite{Enderle05}.

\begin{figure}
\centering
\includegraphics[width=0.49\columnwidth]{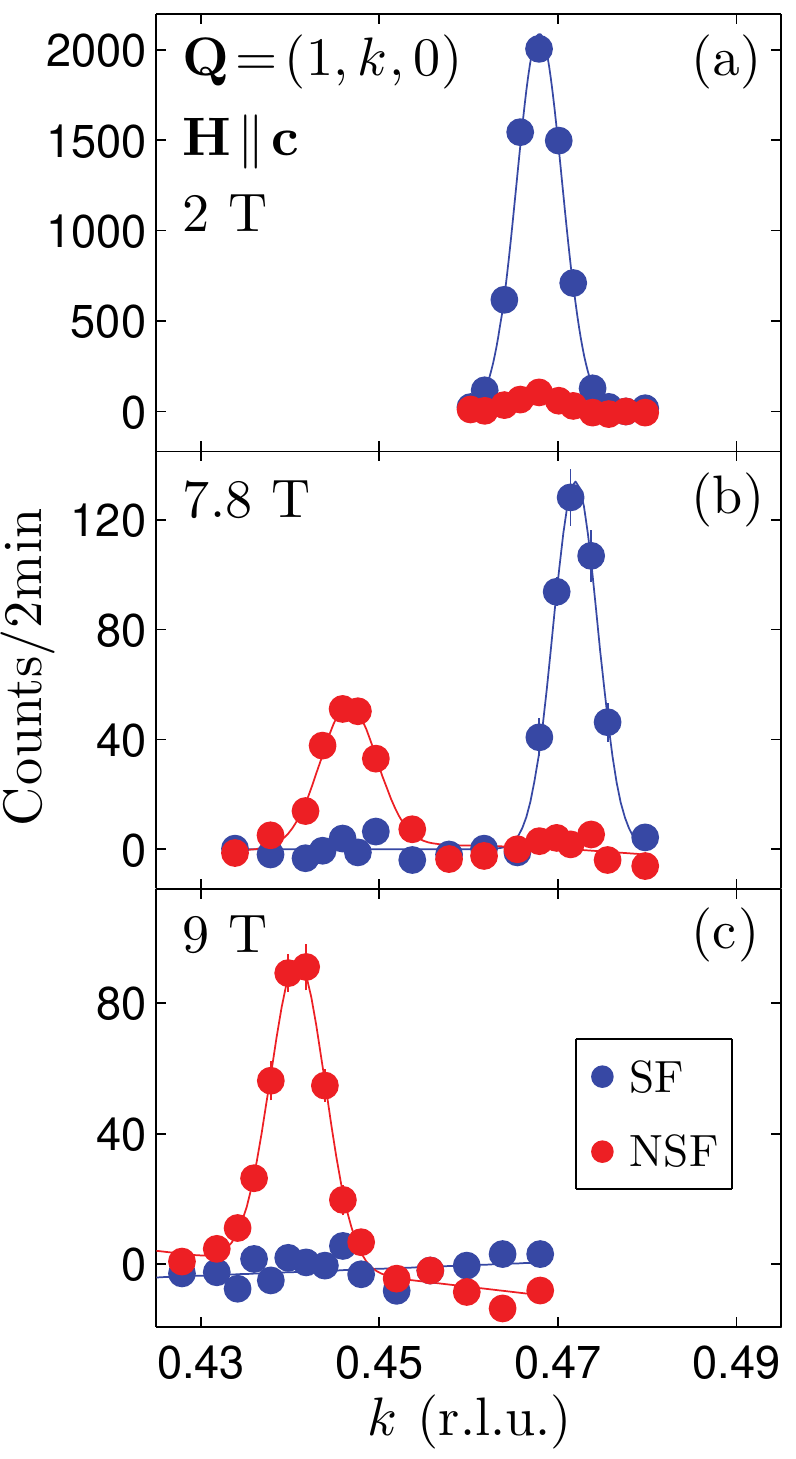}
\includegraphics[width=0.49\columnwidth]{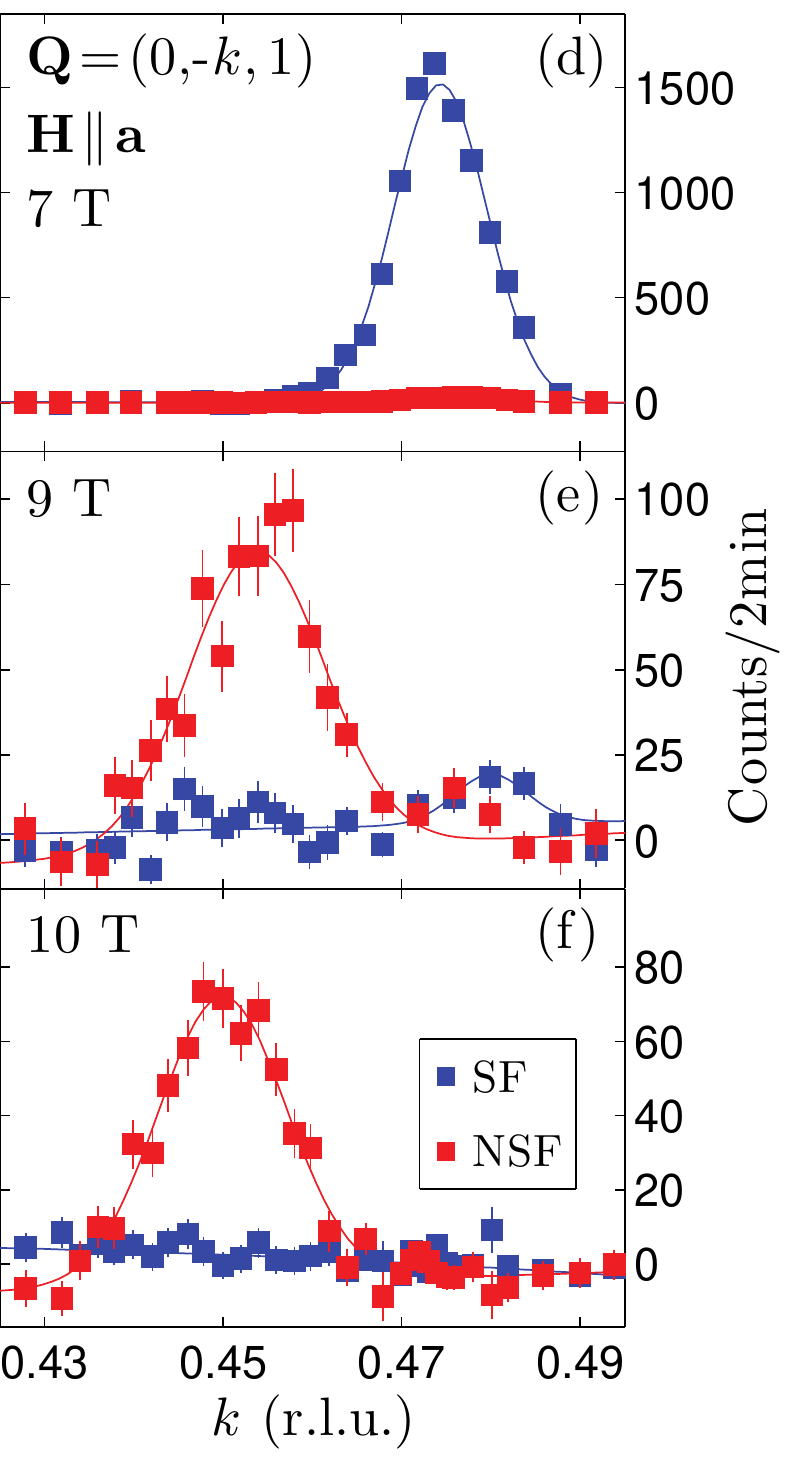}
\caption{(Color online) Polarized cross-sections measured at $T\!=\!70$~mK for the magnetic reflections ${\bf Q}\!=\!(1,\kicy,0)$ with \Hpc\ (left panels, a--c) and ${\bf Q}\!=\!(0,-\kicy,1)$ with \Hpa\ (right panels, d--f).}
\label{FigPA}
\end{figure}

To determine which spin components are involved in the 
SR dipolar correlations above \HQ, we performed a second series of elastic neutron scattering experiments, employing polarized neutrons in a vertical magnetic field with \Hpc\ or \Hpa. The measurements were performed on the cold triple-axis spectrometer IN14 using an asymmetric cryomagnet reaching up to 12~T and a dilution refrigerator going down to 70~mK. Neutrons of wave vector 1.5~\AA$^{-1}$ from a PG$(002)$ monochromator were polarized by a supermirror bender and their polarization, being parallel or anti-parallel to the applied field, was controlled by a spin flipper in the incoming beam and analyzed with the (111) reflection of a Heusler crystal. 
After flipping ratio corrections and subtraction of the spin-incoherent background, the measured  non-spin-flip (NSF) and spin-flip (SF) cross-sections are in this set-up proportional to dipolar correlations polarized parallel to the field (perpendicular to the scattering plane) and perpendicular to the applied field (and also perpendicular to the wavevector transfer {\bf Q}). The results, shown in Fig.~\ref{FigPA}, will now be discussed. 

The magnetic Bragg reflection $(1,\kicy,0)$ measured for \Hpc\ [Fig.~\ref{FigPA}(a--c)] evolves from purely SF character at 2~T to purely NSF character at 9~T with a remarkable coexistence of the two components in the vicinity of \HQ\ at 7.8~T. This shows unambiguously that the dipolar moments swap in a first order transition from the $ab$-plane, transverse to the applied magnetic field, to the $c$ axis, parallel to the applied field. Similarly, the reflection $(0,-\kicy,1)$ measured for \Hpa\ above the 3.5~T spin-flop field [Fig.~\ref{FigPA}(d--f)] reveals that the scattering changes from SF at 7~T (transverse dipolar, now with moments in the $bc$ plane) to NSF at 9 and 10~T with moments along the field direction $a$. 
For both magnetic-field orientations we therefore conclude that the incommensurate dipolar correlations are LR and transverse to the applied field at low fields and become SR and longitudinal for fields above \HQ. The use of neutron scattering allows us to draw such conclusions without an {\it a priori} model.
Since the dipolar correlations become SR above \HQ, the distinct anomaly in $C_p/T$ above \HQ\ cannot result from a dipolar order parameter. The particular field dependence of the characteristic wave vector \kicy\ shown in Fig.~\ref{FigKC}(a,b) implies that the phase transition is driven by quadrupolar correlations. 

\begin{figure}
\includegraphics[width=0.99\columnwidth]{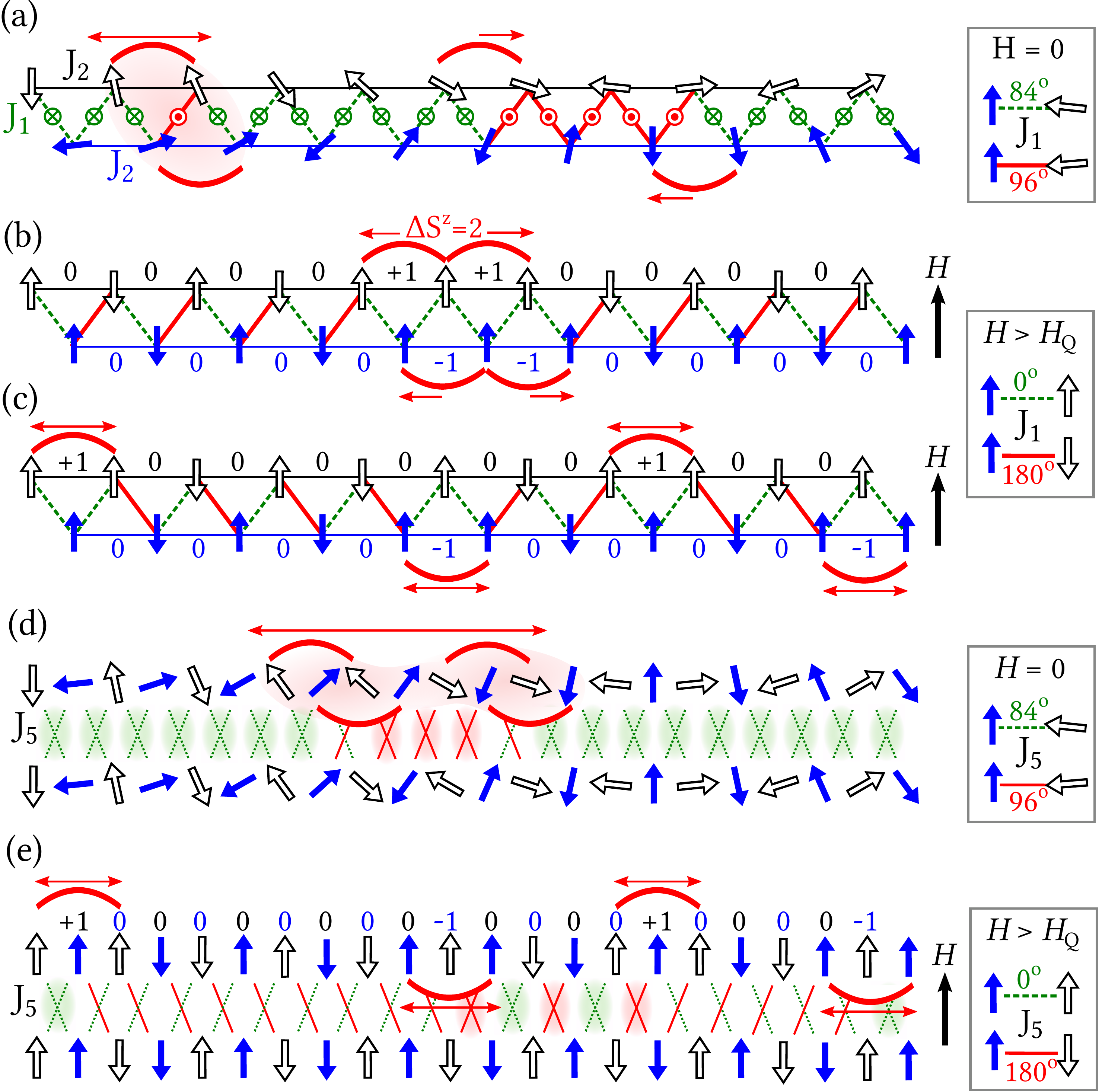}
\caption{ (Color online) (a) FFH-chain at $H\!=\!0$ seen as two intercalated Heisenberg AF chains (closed and open arrows), vertically displaced to illustrate the direction of the NN-bond vector chirality (green crosses and red dots). N\'{e}el solitons (red solid arcs) propagate freely on a Heisenberg AF chain and are bound to $S\!=\!1$ states by $J_1$ in the FFH-chain, resulting in vector-chiral LR order. (b,c) FFH-chain for $H>\HQ$. (b) Two-magnon (four-soliton) states provide $\Delta S^z\!=\!2$ magnetization steps. (c) Propagating individual solitons costs no extra $J_1$-energy, as long as one soliton in the even subchain is followed by one in the odd subchain. This non-local order is reflected in the AF sequence of the bond pseudo-spin $\sigma_n^z$ (numbers). (d,e) Two FFH-chains coupled by a FM diagonal interchain interaction $J_5$. (d) $H\!=\!0$: $J_5$ binds two two-solitons into a four-soliton bound state and thus stabilizes dipolar LR order. (e) $H>\HQ$: Propagation of individual solitons  costs no extra $J_5$-energy, as long as the non-local AF order of the bond pseudo-spin is kept.} 
\label{FigSketch}
\end{figure}

We will now compare our experimental results on \licuvo\ to theoretical predictions. We start with the purely 1D $S=1/2$ FM frustrated Heisenberg (FFH) chain with NN FM exchange $J_1$, AF NNN exchange $J_2$, and $|J_1/J_2|<2.5$ \cite{Heidrich06,Vekua07,Kecke07,Hikihara08,Sudan09}. At zero and small magnetic field \Hpz, uniform LR order of the bond vector-chirality $({\bm S}_n \times {\bm S}_{n+1})^z$ is predicted while transverse and longitudinal dipolar correlations are critical with a field-independent propagation vector close to $k\!=\!1/4$, indicating incipient spiral order \cite{Hikihara08,Sudan09}. The magnetization increases in steps of $\Delta S^z\!=\!1$ \cite{Heidrich06,Kecke07,Hikihara08,Sudan09}. These characteristic features are illustrated using a soliton (fermion) picture in Fig.~\ref{FigSketch}(a), where the magnetization is increased by adding $S\!=\!1$ two-soliton bound states into the ground state.
\HQ\ is identified as a transition to $\Delta S^z\!=\!2$ steps in the magnetization~\cite{Heidrich06,Vekua07,Hikihara08} [Fig.~\ref{FigSketch}(b)]. This is the signature for a two-magnon Bose condensate ground state~\cite{Kecke07,Hikihara08}, with critical transverse bond-quadrupolar correlations  $\left\langle S_0^+ S_{2n+1}^+S_{\ell}^{-}S_{\ell+2m+1}^-\right\rangle$ \cite{Hikihara08}, including correlations of the two-magnon bound state propagator, $\left\langle S_0^+S_{1}^+ S_{\ell}^{-}S_{\ell+1}^-\right\rangle$ \cite{Kecke07,Vekua07,Hikihara08}.  The density fluctuations of the two-magnon bound state condensate \cite{Kecke07,Hikihara08} lead to critical correlations $\left\langle S_n^z S_{m}^z\right\rangle$ \cite{Kecke07,Vekua07,Hikihara08} with a propagation vector $k=1/4-m/2$, characteristic for the whole quadrupolar-nematic two-magnon condensate phase. Transverse dipolar correlations are SR in this phase \cite{Vekua07,Kecke07,Hikihara08}. These characteristics can likewise be illustrated in a soliton image [Fig.~\ref{FigSketch}(c)], where propagation of individual solitons destroys dipolar LR order.
The predictions for the perfect FFH-chain correspond well to the observation of {\it vector-chiral} LR order in \licuvo\ \cite{Mourigal11} with a field-independent \kicy\ at low field,  and the presence of longitudinal dipolar correlations with a propagation vector characteristic of {\it quadrupolar} correlations above \HQ, observed in this work. 

However, \licuvo\ does not correspond exactly to the purely 1D FFH-chain discussed above, since it has a weak easy-plane anisotropy \cite{Mourigal11} and weak interchain interactions \cite{Enderle05}. We now consider these two effects in turn. 
The main effect of exchange anisotropy is to shift \HQ\ to higher fields \cite{Heidrich09}, without any qualitative changes to the vector-chiral and quadrupolar phases in the range of $J_1/J_2$ relevant for \licuvo. However, the large value of \HQ\ of 8 T cannot be reconciled with the very weak easy-plane anisotropy of only 0.5\%~\cite{Mourigal11}. 
The effect of 2D interchain interactions has been studied in the high-field isotropic exchange limit \cite{Zhitomirsky10}. Close to the saturation field \HS, the transverse quadrupolar correlations $\left\langle S_0^+ S_{1}^+S_{\ell}^{-}S_{\ell+1}^-\right\rangle$ are shown to be LR ordered, while (apart from the uniform component $m$) the longitudinal and transverse dipolar correlations are both SR. Below a critical field \HC\ (close to \HS), this LR ordered bond-quadrupolar phase becomes unstable. In \licuvo, \HC\ is found at 40-48~T from magnetization measurements \cite{Svistov11}. Surprisingly, \licuvo\ shows the essential features of this high-field bond-quadrupolar phase already just above \HQ, namely SR dipolar correlations, quadrupolar correlations, and a phase transition evidenced by the specific heat. Our work does not give support for a hypothetical LR longitudinal dipolar (amplitude-modulated) phase in the lower-field region of the quadrupolar-nematic phase \cite{Hikihara08,Heidrich09,Sudan09,Buettgen07,Buettgen12,Masuda11}. 

We now argue that the peculiar type of order observed in \licuvo\ is related to the diagonal nature of the dominant interchain exchanges: the FM $J_5$ acts diagonal in the $ab$ plane, while the much smaller AF $J_6$ acts along the main body diagonal \cite{Enderle05}. Within the soliton picture, at zero magnetic field, $J_5$ and $J_6$ favor bound four-soliton states, and thus spiral dipolar LR order, which is only little perturbed by bound four-soliton propagation [Fig.~\ref{FigSketch}(d)]. This completely agrees with the observed dipolar LR order in \licuvo\ at $H=0$. Moreover, inelastic neutron scattering evidences bound two-spinon (two-soliton) excitations and additional  strong four-spinon excitations \cite{Enderle10}, which may find their explanation in a ground state that already contains bound four-solitons. Above \HQ, in contrast to the zero-field case, neither $J_5$ nor $J_6$ bind the four solitons [Fig.~\ref{FigSketch}(e)], and the propagation of individual solitons destroys longitudinal dipolar LR order in all reciprocal space directions, as we observe experimentally in \licuvo. 

In the soliton picture of the FFH-chain above \HQ, the NNN bond pseudo-spins, $\sigma_n^z=\frac{1}{2}(-1)^n (4 S_n^z S_{n+2}^z +1)$ (whose values are reported in Fig.~\ref{FigSketch}), form a non-local LR AF string order with finite values of the correlation function $G(n)=\langle \sigma_{n_0}^z \exp[i\pi \sum_{m=n_0}^{n+n_0} \sigma_m^z] \sigma_{n+n_0}^z \rangle$ for $n\rightarrow \infty$ \cite{Nijs89}, resembling the string order in valence-bond solid spin-1 chains and the Haldane ground state. This means that for finite FM $J_1$, a propagating pseudo-spin $\sigma_n^z=+1$ cannot go past neighboring $\sigma_m^z=-1$ and vice versa. Since only the direction of the bond-pseudo spin is ordered, but not its position, the AF string order of the bond-pseudo spin is nematic in a sense very close to liquid-crystalline phases. Finite FM $J_5$ or AF $J_6$ not only favor the AF string order along the FFH-chains, but also coherence between chains such that a pseudo-spin $\sigma_n^z=+1$ cannot propagate past $\sigma_m^z=-1$ states in neighboring chains. This leads to SR dipolar correlations in the $ab$-plane and between planes. Since the LR string order breaks at least one discrete $Z_2$ symmetry, with a certain 2D or 3D coherence, it could lead to a finite-temperature phase transition. We speculate that the distinct anomaly in the specific heat above \HQ\ evidences a transition to such a 2D or 3D correlated low-temperature phase with AF alignment of the bond pseudo-spins but no positional order. 
Further theoretical investigations of the order parameter in the quadrupolar-nematic phases of coupled FFH-chains are clearly needed to fully explain our experimental results.

We are grateful to Mike Zhitomirsky for stimulating discussions. The work of M.M. was supported in part by the US Department of Energy, office of Basic Energy Sciences, Division of Material Sciences and Engineering under grant DE-FG02-08ER46544.

\end{document}